\documentclass[preprint,12pt,authoryear]{elsarticle}

\usepackage{titlesec}

\usepackage{etoolbox}

\makeatletter
\def\ps@pprintTitle{%
  \let\@oddhead\@empty
  \let\@evenhead\@empty
  \let\@oddfoot\@empty
  \let\@evenfoot\@oddfoot
}
\makeatother
  
\titleformat{\paragraph}
{\normalfont\normalsize}{\theparagraph}{1em}{}
\titlespacing*{\paragraph}
{0pt}{3.25ex plus 1ex minus .2ex}{1.5ex plus .2ex}

\usepackage{graphicx}
\usepackage{subcaption}
\usepackage[margin=1in]{geometry}  
\usepackage{amssymb}
\usepackage{float}
\usepackage{amsmath}
\usepackage{graphics}
\usepackage{caption}  
\usepackage[utf8]{inputenc}
\usepackage[T1]{fontenc}
\usepackage[russian,english]{babel}

\begin{document}

\begin{frontmatter}



\title{Systemic Decarbonization of Road Freight Transport: A Comprehensive Total Cost of Ownership Model}


\author[label1,*]{Ruixiao Sun}
\author[label1]{Vivek A. Sujan}
\author[label1]{Gurneesh Jatana}

\affiliation[label1]{organization={Oak Ridge National Laboratory},
             addressline={1 Bethel Valley Rd},
             city={Oak Ridge},
             postcode={37930},
             state={TN},
             country={US}}
             

             

             
\affiliation[*]{Corresponding author}
\begin{abstract}

The decarbonization of road freight transport is crucial for reducing greenhouse gas emissions (GHG) and achieving climate neutrality goals. This study develops a comprehensive Total Cost of Ownership (TCO) model to evaluate the economic viability and strategic pathways for decarbonizing road freight transport. The model integrates vehicles with infrastructures, encompassing costs associated with acquisition, operation, maintenance, energy consumption, environmental impacts, and end-of-life considerations. Our analysis covers medium- and heavy-duty vehicles across eight powertrain types, with variants on battery sizes and fuel cell powers, incorporating key financial parameters, technological advancements, and policy incentives. Data sources include real-world fleet data and costs gathered from expert interviews, cross-referenced with multiple public resources. Findings indicate that zero-emission and near-zero-emission vehicles, though currently more expensive, will become cost-competitive with diesel vehicles by leveraging advancements in battery, fuel cell, and hydrogen technologies. 

\end{abstract}




\end{frontmatter}

\section{Introduction}
The transportation sector is the largest source of direct greenhouse gas (GHG) emissions and the second largest when considering indirect emissions from electricity end-use across sectors. It accounted for over 28\% of total direct GHG emissions and 29\% of total indirect GHG emissions in 2022 \citep{EPA2022}. Despite overall reductions in GHG emissions across various transport modes over the past three decades, emissions from heavy-duty vehicles (HDVs) have increased by 83\% since 1990 and by 5\% since 2005. This increase is primarily due to the surge in freight activity and minimal improvements in fuel economy \citep{EPA2022}. Given the sector's rapid growth—road freight activity grew by an average of 1.8\% annually from 1995 to 2018, outpacing the 1.0\% growth in road passenger activity—electrifying HDVs becomes crucial \citep{basma2022fuel}. Without significant intervention, the expected increase in freight activity will likely negate the benefits of CO\textsubscript{2} reduction mandates, leading to a projected 8\% rise in emissions by 2050 \citep{Mulholland2022}.

There are several opportunities to reduce GHG emissions associated with road freight transport as suggested by the U.S. EPA, including: 1) Adopting fuels that produce lower CO\textsubscript{2} emissions compared to those currently in use. Potential alternatives are biofuels, hydrogen, electricity derived from renewable sources such as wind and solar, or fossil fuels with a reduced CO\textsubscript{2} footprint \citep{EPA2023a}. 2) Advancing vehicle technology to include hybrid and electric vehicles capable of capturing braking energy and repurposing it for later use \citep{EPA2023b}. Zero-emission HDVs (ZE-HDVs), including battery-electric vehicles (BEVs) and fuel cell electric vehicles (FCEVs), and alternative fuel engines, i.e., H2-ICE and near-zero-emission HDVs (NZ-HDVs), including hybrid HDVs with diesel (NZEV-D), hydrogen (NZEV-H2), and natural gas (NZEV-NG), show promise in reducing lifecycle GHG emissions. However, market adoption of these technologies is slow due to economic uncertainties and high upfront costs. Over 95\% of global road-freight vehicles still rely on fossil fuels. The International Energy Agency (IEA) forecasts that under a sustainable development pathway, this percentage could drop below 33\% by 2050, replaced by near-zero or zero-emission alternatives \citep{bouckaert2021net}. Achieving this transition necessitates coordinated national and sub-national emissions reduction targets and a shift in vehicle manufacturing priorities \citep{IPCC2022}. Additionally, regulations such as the Advanced Clean Fleet (ACF) mandate a transition to 100\% zero-emissions for on-road vehicles over 8,501 lbs Gross Vehicle Weight Rating (GVWR) by 2042. For commercial fleet agencies and government policymakers, a strategic decarbonized pattern is necessary to guide this transition. However, identifying plausible pathways is challenging due to significant uncertainties in economic and technological developments.

The purpose of this study is to develop a comprehensive Total Cost of Ownership (TCO) model to assess the feasibility and cost-effectiveness of near-zero or zero-emission alternatives in road freight decarbonization. This study also investigates the incremental costs of charging and refueling infrastructures. This developed model aims to answer the following questions:

\begin{enumerate}
\item What are the system TCOs of near-zero or zero-emission alternatives and diesel vehicles, considering current and future infrastructure costs?
\item When will these near-zero or zero-emission alternatives become cost-competitive with their diesel counterparts?
\item What are the key factors impacting the cost of alternative vehicles and their transition viability from diesel?
\end{enumerate}

The literature on the cost of transitioning to sustainable transport primarily adopts two approaches: scenario analysis for policy assessment and comparative studies of TCO for different technologies. Scenario-based analyses model potential future scenarios to evaluate the impacts of various policies on transport systems \citep{parviziomran2023cost, rottoli2021alternative}. These studies aim to assess how different policy interventions might influence the adoption of sustainable transport options, therefore widely used to evaluate vehicle economics, supports informed decision-making by consumers, manufacturers, and policymakers. Nevertheless, existing models often focus on limited vehicle classes and do not fully integrate all relevant cost factors \citep{bjerkan2016incentives, bubeck2016perspectives, hagman2016total, breetz2018electric}. 
Comparative TCO studies, on the other hand, assess the lifetime costs of different technologies to determine the competitive conditions for low-carbon options. These studies are becoming increasingly prevalent as they provide valuable insights into the economic viability of various transport technologies \citep{blanco2019, hagos2018}. However, these studies often lack comprehensive scope, frequently omitting infrastructure costs, future cost projections, detailed lifecycle cost breakdowns, and environmental impact assessments. While many studies include initial purchase price and fuel expenses, fewer account for maintenance, insurance, taxes, subsidies, refueling/charging time, and salvage value \citep{breetz2018electric, parker2021saves, mercure2018integrated, lam2021policy, gnann2017best, alonso2022technical, ahmadi2019environmental, hagos2018, palmer2018total}. Therefore, both scenario analyses and comparative TCO studies have limitations in fully capturing the comprehensive costs and complexities associated with transitioning to sustainable transport. 

Most of the existing models focus on techno-economic analysis of light duty vehicles, recently, several recent studies have extended comparative TCO analysis to medium and heavy-duty vehicles. For instance, \cite{noll2022analyzing} conducted a comprehensive TCO analysis of low-carbon drive technologies in European road freight, comparing battery-electric, hydrogen fuel cell, and diesel trucks. Their findings indicate that battery-electric trucks could become cost-competitive with diesel counterparts under certain conditions, such as decreasing battery costs and supportive infrastructure development. \cite{rajagopal2024comparative} evaluated the TCO of battery-electric versus diesel trucks in India. Their comparative study highlighted that battery-electric trucks have the potential to be economically advantageous, especially with policy interventions that reduce upfront costs and enhance charging infrastructure. 
\cite{basma2022fuel} analyzed the TCO of fuel-cell hydrogen long-haul trucks in Europe. The study assessed the economic feasibility of hydrogen-powered trucks and identified key factors such as fuel prices, infrastructure availability, and technological advancements that influence their competitiveness against diesel trucks. Lastly, \cite{burnham2021comprehensive} provided a comprehensive TCO quantification across different vehicle size classes and powertrains, including medium and heavy-duty vehicles. Their analysis encompassed various technologies like battery-electric and fuel cell vehicles, offering insights into cost dynamics. 

These studies collectively contribute to a deeper understanding of the economic considerations for medium and heavy-duty vehicles transitioning to low-carbon technologies. However, they also reveal limitations in existing models, such as regional specificity, technology focus, and the need for more integrated approaches that consider a broader spectrum of cost factors and operational conditions. 

This study addresses these gaps by developing an integrated TCO model that includes both vehicle and infrastructure costs, incorporates future cost projections, provides detailed cost breakdowns, and includes environmental impact assessments. Our model is structured into three interconnected modules—vehicle, “local” infrastructure, and “regional” infrastructure—to provide a comprehensive evaluation of decarbonization strategies.

The key contributions of this study are:
\begin{itemize}
    \item Development of a comprehensive TCO model encompassing acquisition, operation, maintenance, energy consumption, environmental impacts, and end-of-life considerations.
    \item Integration of infrastructure costs into TCO analysis, providing a more accurate assessment of the economic viability of alternative fuel technologies.
    \item Inclusion of future projections and technological advancements to estimate long-term costs and breakeven points.
    \item Detailed cost breakdowns for various vehicle classes and powertrain types, considering real-world route and usage data.
    \item Scenario and sensitivity analyses to evaluate the impact of key factors on TCO and identify strategic pathways for decarbonization.
\end{itemize}

The paper is organized as follows: Section \ref{methodology} outlines the model design and methodology used in this study. Section \ref{results} presents the analysis results for capital costs and TCO across different vehicle and infrastructure types, including the comprehensive system of system TCO. It also discusses the findings from the sensitivity and scenario analyses. Finally, Section \ref{conclusion} summarizes the key findings, acknowledges the study's limitations, and proposes directions for future research.

\section{Model Descriptions} \label{methodology}
To evaluate the economic viability and strategic pathways for decarbonizing road freight transport, considering alternative fuel technologies, infrastructure requirements for charging and refueling, and the integration of power grid support with renewable energy generation, we developed a comprehensive model that synthesizes various methods, as illustrated in Fig. \ref{fig:overview}.

\subsection{Model overview}

The model is structured into two discrete but interconnected modules—vehicle, “local” infrastructure, and “regional” infrastructure—each encapsulating distinct elements essential to the comprehensive evaluation of decarbonization strategies. 
\begin{itemize}

  \item Vehicle module. The vehicle module is segmented by three medium- and heavy-duty vehicle classes, including box truck, day cab, and sleeper configurations, and spans eight powertrain types (diesel internal combustion engine vehicle (D-ICE), hydrogen internal combustion engine vehicle (H2-ICE), natural gas internal combustion engine vehicle (NG-ICE), battery electric vehicle (BEV), fuel cell electric vehicle (FCEV), near-zero electric vehicle with hydrogen (NZEV-H2), near-zero electric vehicle with natural gas (NZEV-NG), and near-zero electric vehicle with diesel (NZEV-D)). 
  \item	"Local" infrastructure module. This module captures the costs associated with various refueling and recharging modalities, from the installation and operation of diesel, hydrogen, and natural gas refueling infrastructures to the intricate demands of electric vehicle charging stations. 
  \item	“Regional” infrastructure module. This module primely evaluates the costs related to energy provision through the power grid, supplemented by DERs such as wind, solar, hydro, and nuclear power. It considers the capital investments required for grid upgrades and the integration of renewable energy sources, offering a future-proof perspective on energy supply costs. It also covers the six cost breakdowns.

\end{itemize}

\begin{figure}[htb] \centering
    \includegraphics[width=\columnwidth]{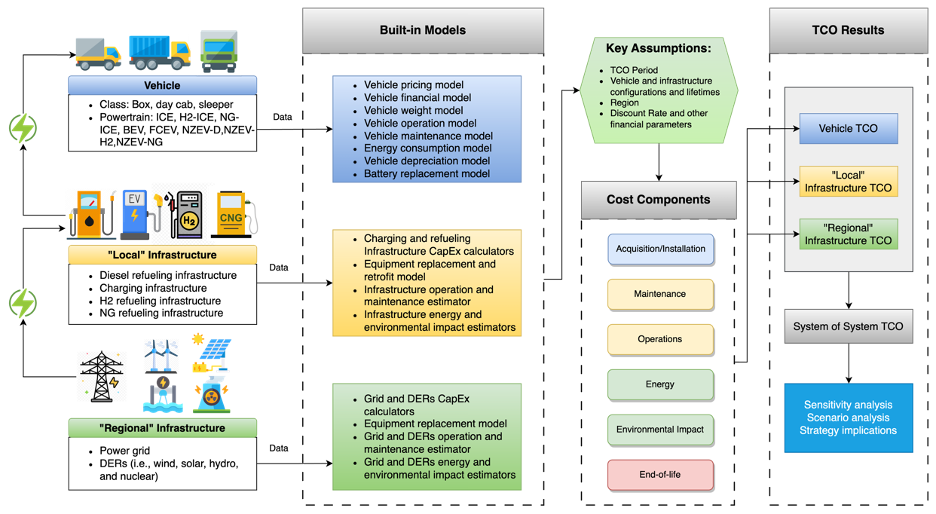}
     \caption{Overview of the developed Total Cost of Ownership (TCO) tool}
     \label{fig:overview}
\end{figure}

As illustrated in Fig. \ref{fig:overview}, the developed TCO model defines and categorizes six primary cost components: acquisition/installation, maintenance, operation, energy consumption, end-of-life considerations, and environmental impacts. Each component is calculated to reflect both direct and indirect expenditures, incorporating detailed data to provide a nuanced understanding of cost implications across different vehicle classes and powertrain types. Each module within the model includes several built-in sub-models to support and execute the estimations and forward projections of these cost components.

For example, the "vehicle module" encompasses a vehicle pricing model and a vehicle financial model to determine acquisition costs. The vehicle weight model, combined with the vehicle operation model, estimates operational costs. The battery replacement model predicts necessary battery replacements based on battery throughput and estimates replacement costs in conjunction with the vehicle maintenance model, which accounts for age and mileage. The energy consumption model simulates fuel or electricity consumption for specific vehicles based on real driving routes and patterns. Additionally, the vehicle depreciation model simulates depreciation rates over time, aiding in the determination of end-of-life costs.

The "local infrastructure module" includes capital expenditure (CapEx) calculators for charging infrastructure and three types of fuel refueling infrastructures, providing detailed calculations for construction and installation capital costs. It also incorporates an equipment replacement and retrofit model to account for replacement costs. Infrastructure operation and maintenance estimators are proposed to assess operation and maintenance expenses, while infrastructure energy and environmental impact estimators evaluate energy costs and environmental impact costs. This module also features an infrastructure depreciation model to calculate end-of-life costs for depreciable equipment. Similarly, the "regional infrastructure module" includes CapEx calculators for the power grid and DERs, an equipment replacement model, grid and DERs operation and maintenance estimators, and grid and DERs energy and environmental impact estimators to comprehensively calculate these cost components.

The developed TCO model is designed to analyze costs over a long-term horizon while offering customization options for shorter periods. It comprehensively captures both direct and indirect costs, allowing for flexible customization of vehicle and infrastructure configurations, region-specific cost adjustments, and financial parameters. The model outputs lifetime and levelized costs for vehicles by class and powertrain type, as well as for infrastructures by type. By interconnecting and integrating the three modules into a unified layer, it generates a system of system TCO, representing the combined cost for each vehicle type. This facilitates comparisons of alternative fuel technologies with their diesel counterparts from a holistic system perspective. Furthermore, the model supports both sensitivity and scenario analysis, enabling the examination of key parameters' influence on the TCO and the comparison of different strategic scenarios for vehicle and infrastructure investments. This capability provides valuable insights for strategic pathway implications.

Notably, this study focuses on the vehicle module and the "local" infrastructure module, which have been fully developed within our proposed framework. The "regional" infrastructure module is currently under development and will be demonstrated in future studies. In the following sections, we will continue to use the term "infrastructure" to refer specifically to the "local" infrastructure.

\subsection{Model details}

This section demonstrates the model details within the scope of vehicles and infrastructures. The TCO is computed over a customizable analysis period using the following equation:

\begin{equation} \label{eqn:1}
    TCO = \sum_{i=1}^N\frac{C_i}{(1+d)^i}
\end{equation}

\noindent where $N$ is the analysis period, $d$ is the discount rate, and $C_i$ represents a cash flow of all the costs in year $i$, in real dollars, adjusted for inflation, but not discounted.

To offer a granular economic perspective, costs are further amortized over traveled distance, producing levelized costs on a per-mile basis by employing the equation proposed in \citep{burnham2021comprehensive}.

\begin{equation} \label{eqn:2}
    LC = \sum_{j}LC_j = \sum_{j}\frac{TCO_j}{\sum_{i=1}^N\frac{VMT_i}{(1+d)^i}}
\end{equation}

\noindent where $VMT$ represents vehicle miles traveled and $j$ indexes the individual cost components.

\subsubsection{Capital Cost}
Capital cost plays a critical role in TCO analysis. This section describes the detailed calculation of capital costs for both vehicles and infrastructures.

\paragraph{Vehicle Capital Cost}

As aforementioned, the "Vehicle Module" includes a vehicle pricing model to compute prices for specific vehicle classes and powertrain types. We divide the vehicle price into several sub-parameters based on key components, including chassis, 12-volt system, air compressor, heating, ventilation, and air conditioning (HVAC), power steering, engine, transmission and clutch, electric traction drive, high voltage harness, high voltage junction box, onboard charger, hybrid generator set (genset), after-treatment, cooling system, fuel cell, battery, fuel tank, and other vehicle components associated with the vehicle glider, such as the body shell, wheels, cabin, etc. We apply multipliers to represent the scales and differences among different vehicle classes (e.g., box truck, day cab, and sleeper) and powertrain types. These multipliers have been validated and calibrated with OEM-suggested data. The price of sub-components can be expressed as,

\begin{equation} \label{eqn:3}
    P_{jkh} = C_{0,0,h} \times F_{jkh} \times M_{jk}
\end{equation}

\noindent where $C_{0,0,h}$ is the manufacturing cost of the baseline (D-ICE box truck) on the sub-component $h$. $F_{jkh}$ represent the profit margin for component $h$, class $j$, and powertrain $k$. $M_{jk}$ represent the multipliers for class $j$ and powertrain $k$.

In addition, the model accounts for the purchasing price of powertrains over time, influenced by technological advancements due to learning by doing \citep{sofia2020cost}. The learning rate, which is defined as the cost reduction achieved through the growth in cumulative installed capacity \citep{karkatsoulis2017simulating}, plays a crucial role in capital costs.  For each sub-parameter, we apply a learning curve to represent anticipated technological advancements. Additionally, we consider tax incentives for clean vehicles. The adjusted vehicle price accounting for the estimated purchase price with the subtraction of incentives can be expressed as:

\begin{equation} \label{eqn:3}
    VP_{i,jk} = \sum_{k}P_{i,jkh} - I_{i,jk}
\end{equation}

\noindent where $P_{i,jkh} =P_{0,jkh} \times (1+r_{jkh}^L)^n$ denotes the future price of vehicle class $j$, powertrain $k$ and vehicle sub-component $h$ at year $i$, considering the learning rate $r_{jkh}^L$. $I_{i,jk}$ represents the incentives of vehicle class $j$ and powertrain $k$ at year $i$. 

The financial terms are modeled in the vehicle financial model to capture nuanced differences in down payments, interest rates, and finance terms across different vehicle types and powertrains. The vehicle capital cost can be computed by,

\begin{equation} \label{eqn:5}
    \text{CapEx}^{Vehicle} = VP \times r_D + \sum_{i=1}^N\frac{VP \times (1-r_D) \times  \frac{r \times (1 + r)^i}{(1 + r)^i - 1}}{(1+d)^i}
\end{equation}

\noindent where $r$ is the interest rate, $r_D$ is the ratio of down payment and $i$ is the year.

\paragraph{Infrastructure Capital Cost}

The model is equipped with CapEx calculators for diesel refueling, charging, hydrogen refueling, and natural gas refueling infrastructures to supply the four types of energy for the eight powertrain types. The calculators account for both equipment costs and development costs. Equipment costs include detailed components specific to each type of infrastructure. For instance, charging infrastructure requires chargers, on-site utility equipment, and various software, control, and safety equipment. Hydrogen refueling infrastructure includes a dispenser, compressor, cascade storage, electrical equipment, refrigeration system, and associated software, control, and safety equipment. Natural gas refueling infrastructure is equipped with a dispenser, compressor, storage, electrical equipment, dryer system, and related software, control, and safety equipment. Diesel refueling infrastructure comprises a pump, compressor, storage, electrical equipment, and relevant software, control, and safety equipment. Development costs cover fieldwork for site construction and equipment installation, as well as office work such as permitting, engineering and design, project management, and contingency planning.

The calculators generate detailed breakdowns of capital costs based on the given infrastructure configuration, including the number of dispensers/chargers, filling rate, and utilization rate. Additionally, the model accounts for technological advancements that reduce equipment costs, represented by the learning rate. The infrastructure capital cost for a given infrastructure configuration in year $i$ can be calculated as follows:

\begin{equation} \label{eqn:5}
\text{CapEx}^{\text{Infrastructure}}_i = \sum_{l}^m A_l E_{il} + \sum_{f}^n B_f D_{if}
\end{equation}

\noindent where $A_l$ and $B_f$ are the quantities of equipment and development items needed for the type of infrastructure, respectively. $E_{il} = E_{0,l} \times (1 + r_{l}^L)^n - I_{il}$ and $D_{if} = D_{0,f} \times (1 + r_{f}^L)^n$ represent the unit costs of equipment and development items in year $i$, considering the effects of technology learning by applying the learning rate $r_{l}^L$ for equipment and $r_{f}^L$ for development.

\subsubsection{Maintenance Cost}

Maintenance is a significant ongoing expense in the TCO for vehicles and infrastructure in the freight transport sector. Maintenance costs for vehicles are categorized into scheduled and unscheduled services, repairs, replacements, labor, and adjustments based on warranty coverage \citep{liu2021comparing}. These costs are highly dependent on vehicle miles traveled (VMT) and vary across different types of vehicles. Additionally, key vehicle component replacement costs, such as battery replacement, are considered.

\begin{equation} \label{eqn:5}
\text{Maintenance}^\text{Vehicle}_i = 
    \begin{cases}
      F_{jk}(M_i) \times \text{VMT}_i, & \text{if } L_{ih} < \text{life}_h \\
      F_{jk}(M_i) \times \text{VMT}_i + P_{ih} \times (1 + r_\text{labor}), & \text{if } L_{ih} \geq \text{life}_h
    \end{cases}
\end{equation}

\noindent where $F_{jk}(M_i)$ is the unit maintenance cost for class $j$ and powertrain $k$ based on the vehicle odometer mileage $M$ in year $i$. $P_{ih} \times (1 + r_\text{labor})$ is the replacement cost, including labor, when the equipment life $L_{ih}$ is greater than or equal to the designed equipment life $\text{life}_h$.

Similarly, infrastructure maintenance includes regular upkeep and emergency repairs for various types of infrastructure, considering the replacement costs of key equipment.

\begin{equation} \label{eqn:5}
\text{Maintenance}^\text{Infrastructure}_i = 
    \begin{cases}
      F_{g}(i) , & \text{if } L_{ih} < \text{life}_h \\
      F_{g}(i) + P_{ih} \times (1 + r_\text{labor}), & \text{if } L_{ih} \geq \text{life}_h
    \end{cases}
\end{equation}
\noindent where $F_{g}(i)$ is the unit maintenance cost for type $g$  based on the infrastructure age in year $i$. 

\subsubsection{Operation Cost}

Operation costs encapsulate all expenses associated with the day-to-day functioning of vehicles and infrastructure. 

For vehicles, the operational costs are multifaceted, depending on the vehicle type and their respective operational characteristics. This includes direct costs such as driver wages, insurance, taxes, fees, and indirect costs like downtime and payload capacity loss.
\begin{equation}
    OpEx^{Vehicle}_i = Driver_i + Insurance_i + Tax_i + Payload^{Loss}_i + Dwell_i
\end{equation}

Operational costs for infrastructure are similarly categorized but tailored to the specifics of charging stations, hydrogen refueling stations, natural gas refueling, including labor costs, insurance, property taxes, expected usage patterns, downtime estimates,  and various fees and licenses.

\begin{equation}
    OpEx^{Infrastructure}_i =  Insurance_i + Warranty_1 + Tax_i + Labor_i + Downtime_i
\end{equation}

\subsubsection{Energy Cost} \label{energy cost}

In-use energy consumption is a critical aspect of the TCO for various vehicle powertrains and infrastructures.

In the vehicle module, the energy cost component considers the type of fuel (diesel, electricity, hydrogen, natural gas) and includes a sophisticated energy consumption simulation model that accounts for both route and vehicle characteristics. Route characteristics describe how the vehicle travels to complete the mission and include variables such as distance traveled, speed limit, elevation, payload carried, and ambient conditions. Given the origin and destination of a mission, online mapping services such as Google Maps or Nokia Here Maps can be used to generate the route a vehicle would travel, including the speed limit and elevation data. Once a route is defined, the energy consumption of various powertrain architectures is assessed using detailed vehicle models provided by OEM experts. These models offer realistic simulations for various vehicle characteristics such as powertrain architecture, efficiency maps for various components, powertrain control behavior, and auxiliary loads. While the specific details of these models are proprietary and cannot be shared, they generally operate similarly to other vehicle simulation models, such as the open-source Autonomie models \citep{halbach2010model}.

The energy cost for vehicles can be calculated based on the energy consumption of electricity and various fuels (e.g., diesel, hydrogen, and natural gas) as follows:

\begin{equation}
     \text{Energy}_i^{\text{vehicle}} = P_i^{\text{fuel}} \times E_i^{\text{fuel}} + P_i^{\text{electricity}} \times E_i^{\text{electricity}}
\end{equation}

\noindent where \( P_{i}^{\text{fuel}} \) and \( E_{i}^{\text{fuel}} \) represent the price per unit and energy consumption of fuel in year \( i \), \( P_{i}^{\text{electricity}} \) and \( E_{i}^{\text{electricity}} \) represent the price per unit and energy consumption of electricity in year \( i \).

Energy costs for infrastructure are modeled to account for the electricity consumption required to operate refueling and charging stations, as well as other energy losses during the dispensing or discharging process. To avoid double-counting energy costs for vehicles and infrastructure, the energy dispensed to vehicles from on-site infrastructure is excluded from the infrastructure energy cost, as it is already accounted for in the vehicle energy consumption.

As denoted in Eq.(12), The electricity cost for infrastructure in year $i$ is modeled as the estimated utility bill, which is a sum of basic charge $C_i^{\text{fix}}$, delivery charge $C_i^{\text{delivery}}$, and transmission charge $C_i^{\text{transmission}}$, adjusted by the demand charge $C_i^{\text{demand}}$. The demand charge is a fee based on the highest level of power drawn during a billing period, intended to cover the utility's costs of maintaining the capacity to meet peak demand \citep{energy2020demand}. $P_i^{\text{electricity}}$ is the average electricity price in year $i$, and $E_i$ is the electricity usage in year $i$.

\begin{equation}
     \text{Utility}_i^{\text{infrastructure}} = C_i^{\text{fix}} + P_i^{\text{electricity}} \times E_i + C_i^{\text{demand}} + C_i^{\text{delivery}} + C_i^{\text{transmission}}
\end{equation}


\subsubsection{Environmental Impact Cost}
The model monetizes environmental impacts by assigning costs to emissions, providing a financial quantification of the freight system's ecological footprint. The environmental costs for vehicles and infrastructure reflect the carbon footprint of their operations, focusing on CO\textsubscript{2} emissions, which can be estimated from energy consumption. As shown in Eq. (11), the emission factor \( \text{Factor}_i^{\text{carbon}} \) for energy type \( i \) is employed to estimate CO2 emissions based on the energy usage \( \text{Energy}_i \) on a tank-to-wheel basis. The monetized value of the CO2 emissions is then used to calculate the environmental impact cost.

\begin{equation}
     \text{Cost}_{\text{carbon}} = \text{Factor}_i^{\text{carbon}} \times \text{Energy}_i \times \text{Price}_{\text{carbon}}
\end{equation}

\subsubsection{End-of-life Cost}

End-of-life costs encompass the culmination of the economic lifespan of both vehicles and infrastructures, factoring in residual market value, potential salvage value, and any costs associated with disposal and documentation \citep{cox2020life}.

Vehicle depreciation is calculated using a regression model proposed by \citep{burnham2021comprehensive}, which considers both age and mileage to accurately estimate residual values, as expressed in the following equation:

\begin{equation}
    \text{RV}(a,m) = C \cdot \exp(A \cdot a + M \cdot m)
\end{equation}

\noindent where $C$ is the regression-estimated retail price at age 0 with no mileage, $a$ is the age in years, $m$ is the mileage in thousands, $\exp(A)$ is the percentage price retention from the previous year, and $\exp(M)$ is the percentage price retention from the previous 1000 miles.

The end-of-life costs for infrastructure also consider the remaining value of depreciable equipment at the end of their operational life or the analysis period.


\subsection{Data Sources}

Table \ref{table:assumptions} presents the key input parameters for the TCO model, encompassing vehicle and infrastructure costs in the context of transport decarbonization. It outlines the initial configurations and acquisition expenses, including the specific attributes of vehicles and the operational metrics of infrastructure. For ongoing financial obligations, it includes the vehicle's operational costs and the infrastructure's maintenance requirements. The model encapsulates broader economic factors like energy costs and end-of-life valuations, alongside the financial parameters of acquisition, interest rates, and loan or lease terms. These assumptions collectively enable a robust evaluation of the financial and environmental implications of decarbonized transport solutions.\\

\begin{table}[h] 
\centering
\caption{Key Parameters for Vehicle and Infrastructure Modules}
\begin{tabular}{|l|p{6cm}|p{6cm}|}
\hline
\textbf{Key Parameters} & \textbf{Vehicle} & \textbf{Infrastructure} \\
\hline
Acquisition & Class, powertrain type, size, weight, and cost for each component (e.g., fuel tank, battery, fuel cell), profit margin, incentives & Number of dispensers, filling rate, utilization rate, Unit cost for each piece of equipment, civil/structural cost, engineering/design, permit, project management \\
\hline
Operation & VMT, registration fee, tax, driver wages, cargo weight, payload capacity loss cost rate, depot refueling/charging ratio, on-shift queuing time, dwell time labor ratio & Insurance, warranty, tax, licensing fee, communication fee, labor cost, downtime cost \\
\hline
Maintenance & Scheduled/unscheduled maintenance, repairs, replacement & Scheduled/unscheduled maintenance, repairs, replacement \\
\hline
Energy & Energy price, energy consumption & Energy price, base charge, demand charge, electricity consumption, energy transfer efficiency \\
\hline
Environmental impact & CO2 factor, emission penalty & Carbon intensity, emission penalty \\
\hline
End-of-life & Vehicle depreciation rate, operational life & Equipment depreciation rate, operational life \\
\hline
\end{tabular}
\label{table:assumptions}
\end{table}

To ensure our model captures technology and economic trends with plausible accuracy, we cross-reference data from multiple sources. These include publicly available data (e.g., U.S. Energy Information Administration (EIA) \citep{decarolis2023annual}, BloombergNEF \citep{trends2023bloombergnef}, market values published on websites), research values from reports and literature, open-source models (e.g., Autonomie \citep{halbach2010model}, HDRSAM \citep{elgowainy2018hydrogen}, EVI-FAST \citep{wood2023evi}), and expert interviews from industry, universities, and national laboratories, as listed in Table \ref{table:interviewees}. All interviews were conducted under the “Chatham House Rule” \citep{rule2020chatham}, and hence no references to interviewees or their affiliations are made.\\

It is important to note that the projected energy prices for diesel, electricity, and natural gas are sourced from the EIA Annual Energy Outlook \citep{EIA2023}. The hydrogen price is interpolated from anticipated target year prices in a DOE Hydrogen and Fuel Cells Program Record \citep{DOE2018}. Additionally, California's projected carbon prices, as reported by BloombergNEF \citep{BNEF2024}, are used to monetize the environmental impact costs.

\begin{table}[h]
\centering
\caption{List of Organizations, Expertise, and Interviewee Roles}
\begin{tabular}{|c|p{5cm}|p{5cm}|p{5cm}|}
\hline
\textbf{\#} & \textbf{Organization} & \textbf{Expertise} & \textbf{Interviewee’s Role(s)} \\
\hline
1 & Commercial Fleet & Transport and supply chain & Director of Energy Transformation \\
\hline
2 & OEM\slash Tier 1 & Powertrain research and development & Senior Technical Manager \\
\hline
3 & OEM\slash Tier 1 & Vehicle charging & Senior Project Manager \& Business Developer  \\
\hline
4 & Carrier & Transportation service & Senior Advisor Sustainable Mobility \\
\hline
5 & Consultancy & Energy technology development and deployment & Senior Technical Leader \\
\hline
6 & Oil and Gas & Refueling infrastructure & Senior Director of Engineering \\
\hline
7 & Utility & Improve travel and develop the business and farming &  Project Manager \\
\hline
8 & University & Vehicle engineering and modeling & Researcher \\
\hline
9 & University & Sustainable infrastructure & Researcher \\
\hline
10 & National Lab & Carbon-free mobility & Researcher \\
\hline
11 & National Lab & Policy-relevant econometric analysis & Research Analyst  \\
\hline
12 & National Lab & Mobility infrastructure and impacts analysis & Research Analyst \\
\hline
\end{tabular}

\label{table:interviewees}
\end{table}

\section{Results and Discussions} \label{results}

\subsection{Capital Costs Results} 
This section calculates the capital cost for vehicles across eight powertrain types and infrastructures for four energy types.

\subsubsection{Capital Costs of Vehicles}

\begin{table}[hbt!]
\centering
\caption{Vehicle Specifications by Class and Powertrain Types}
\begin{tabular}{|l|l|p{3cm}|p{3cm}|p{3cm}|}
\hline
\textbf{Powertrain} & \textbf{Class} & \textbf{Battery size (kWh)} & \textbf{Fuel tank (gal or kg)*} & \textbf{Fuel Cell Power (kW)} \\ \hline
Diesel & Box truck & 0 & 35 & 0 \\ \hline
H2-ICE & Box truck & 0 & 30 & 0 \\ \hline
NG-ICE & Box truck & 0 & 70 & 0 \\ \hline
BEV & Box truck & 300 & 0 & 0 \\ \hline
FCEV & Box truck & 5 & 30 & 150 \\ \hline
NZEV-H2 & Box truck & 100 & 15 & 0 \\ \hline
NZEV-NG & Box truck & 100 & 35 & 0 \\ \hline
NZEV-D & Box truck & 100 & 18 & 0 \\ \hline
Diesel & Day cab & 0 & 240 & 0 \\ \hline
H2-ICE & Day cab & 0 & 70 & 0 \\ \hline
NG-ICE & Day cab & 0 & 300 & 0 \\ \hline
BEV & Day cab & 700 & 0 & 0 \\ \hline
FCEV & Day cab & 50 & 70 & 200 \\ \hline
NZEV-H2 & Day cab & 250 & 47 & 0 \\ \hline
NZEV-NG & Day cab & 250 & 150 & 0 \\ \hline
NZEV-D & Day cab & 250 & 120 & 0 \\ \hline
Diesel & Sleeper & 0 & 240 & 0 \\ \hline
H2-ICE & Sleeper & 0 & 100 & 0 \\ \hline
NG-ICE & Sleeper & 0 & 496 & 0 \\ \hline
BEV & Sleeper & 1000 & 0 & 0 \\ \hline
FCEV & Sleeper & 50 & 100 & 300 \\ \hline
NZEV-H2 & Sleeper & 250 & 67 & 0 \\ \hline
NZEV-NG & Sleeper & 250 & 248 & 0 \\ \hline
NZEV-D & Sleeper & 250 & 120 & 0 \\ \hline
\end{tabular}
\label{table:powertrain_specs}
\begin{flushleft}
\textit{} * Fuel tank (gal or kg) refers to gallons for diesel and kilograms for hydrogen and natural gas.
\end{flushleft}
\end{table}

The primary inputs for calculating vehicle capital costs are the price and size of key zero-emission powertrain components: the battery, fuel cell, hydrogen tank, and electric drive train. These components' costs can vary significantly based on various factors.

Although lithium iron phosphate (LFP) is the most prevalent battery technology for electric vehicles (EVs) globally \citep{wang2022recycling}, this paper does not differentiate between battery chemistries due to the uncertainty of predominant battery chemistry usage in the United States. Adopting a chemistry-agnostic approach, a recent ICCT study estimates that pack-level battery costs for light-duty vehicles (LDVs) in the U.S. will decrease from \$131/kWh in 2022 to \$105/kWh in 2025, \$74/kWh in 2030, and \$63/kWh in 2035 (in 2020 dollars) \cite{slowik2022assessment}. BloombergNEF analysts predict that prices will fall below \$100/kWh by 2026 due to the expansion of raw material extraction, improvements in manufacturing processes, and increased capacity across the supply chain \citep{BNEF2020}.

Battery costs are higher for heavy-duty vehicles (HDVs) than for LDVs due to several factors \citep{Calstart2022}, a) Higher energy density requirements to prevent payload loss, b) Longer durability requirements to accommodate high annual mileages, c) Harsher operating conditions, including extreme temperatures and reduced vibration dampening,
and d) Lower production volumes and poorer economies of scale.

In this study, we use an expert-recommended battery cost of \$250/kWh, which aligns closely with a recent ICCT report's estimates \citep{xie2023purchase}. We also incorporate a range of battery costs from \$140/kWh to \$380/kWh from various sources \citep{Ricardo2021, slowik2022assessment, burnham2021comprehensive, hunter2021spatial, burke2020technology, Calstart2022, noll2022analyzing, phadke2021regional} to calculate the capital costs.

Similarly, for fuel cells, we use an expert-recommended baseline cost of \$300/kW, and a literature-based range from \$190/kW to \$2,200/kW \citep{Berger2020, Ricardo2021, burnham2021comprehensive, hunter2021spatial, burke2020technology, noll2022analyzing} to account for cost variability.

The estimated cost of hydrogen fuel tanks is \$1,000 per kg of usable hydrogen in 2023, with collected costs ranging between \$546/kg \citep{burke2020technology} and \$1,723/kg \citep{Ricardo2021}. These costs are expected to decline to an average of \$675/kg by 2040 \citep{xie2023purchase}.

For electric drive units, the estimated cost in 2023 is around \$70 per kW of continuous nominal power, projected to drop to \$18/kW by 2040 \citep{xie2023purchase}. Ricardo Strategic Consulting \citep{Ricardo2021} shows the steepest learning curve, with costs decreasing from \$90/kW in 2020 to about \$20/kW in 2030. \cite{burke2020technology} report the lowest 2020 costs at \$49/kW, predicting the highest 2030 costs at \$33/kW.

Table \ref{table:powertrain_specs} presents the vehicle specifications on battery size, fuel tank size, and fuel cell power. These specifications and sizes consider the vehicle’s energy efficiency or fuel economy and the average daily mileage of representative use cases provided by one of our interviewed experts.

\begin{figure}[htb] \centering   \includegraphics[width=\columnwidth]{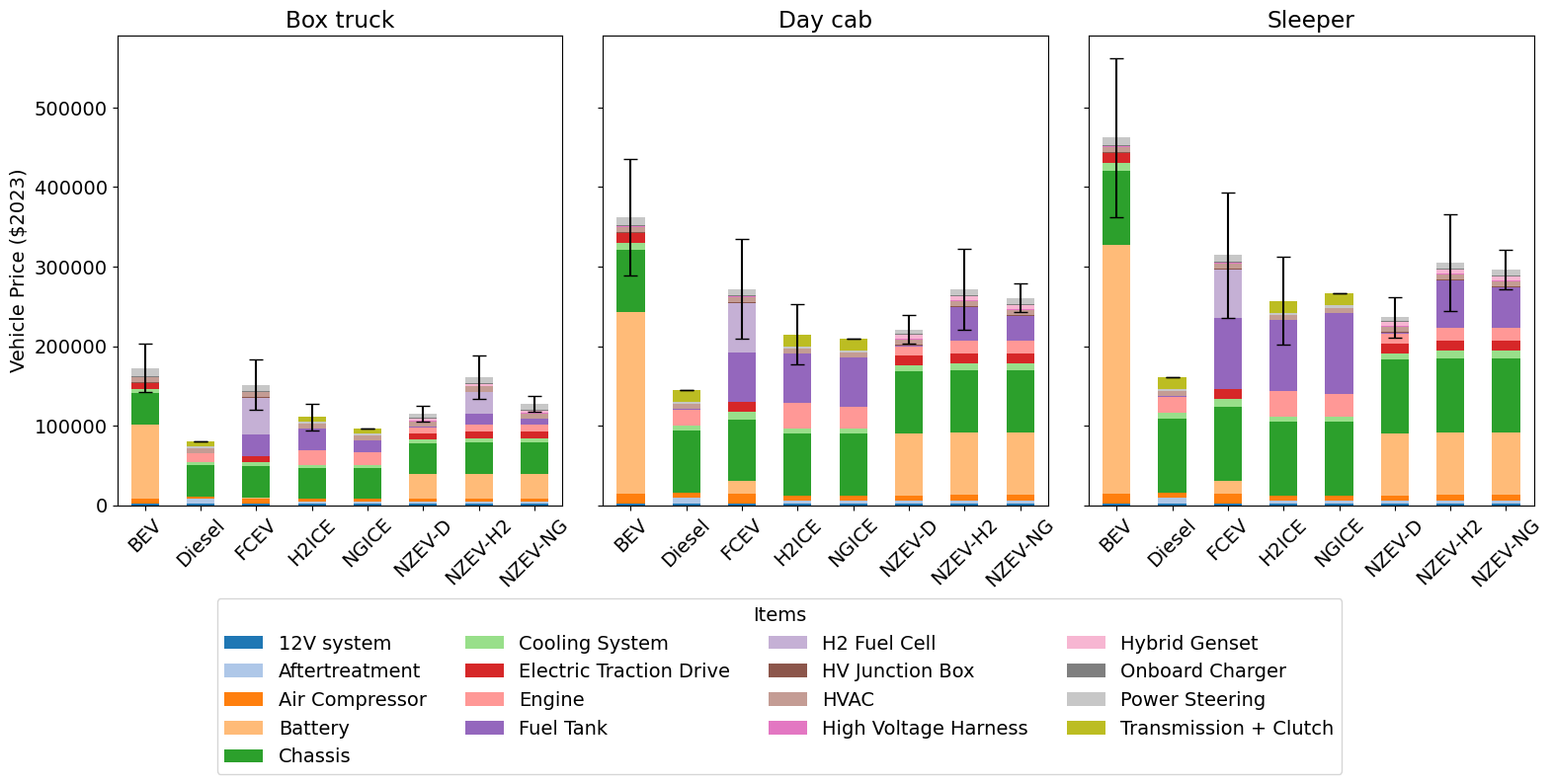}
     \caption{Vehicle price by three classes and eight powertrain types}
     \label{fig:vehicle_price}
\end{figure}

The vehicle price directly determines the capital cost, assuming certain financial parameters. When a fleet company purchases vehicles outright rather than through loans, the vehicle price is considered the capital cost. Figure \ref{fig:vehicle_price} presents the estimated vehicle prices with a detailed breakdown of the vehicle components by three classes and eight powertrain types. This breakdown includes the key inputs on vehicle component size and cost, along with considerations for profit margins. The error bars represent the impact of the ranges of battery, fuel cell, hydrogen tank, and electric drive train costs on the vehicle capital costs.

\begin{figure}[htb] \centering   \includegraphics[width=\columnwidth]{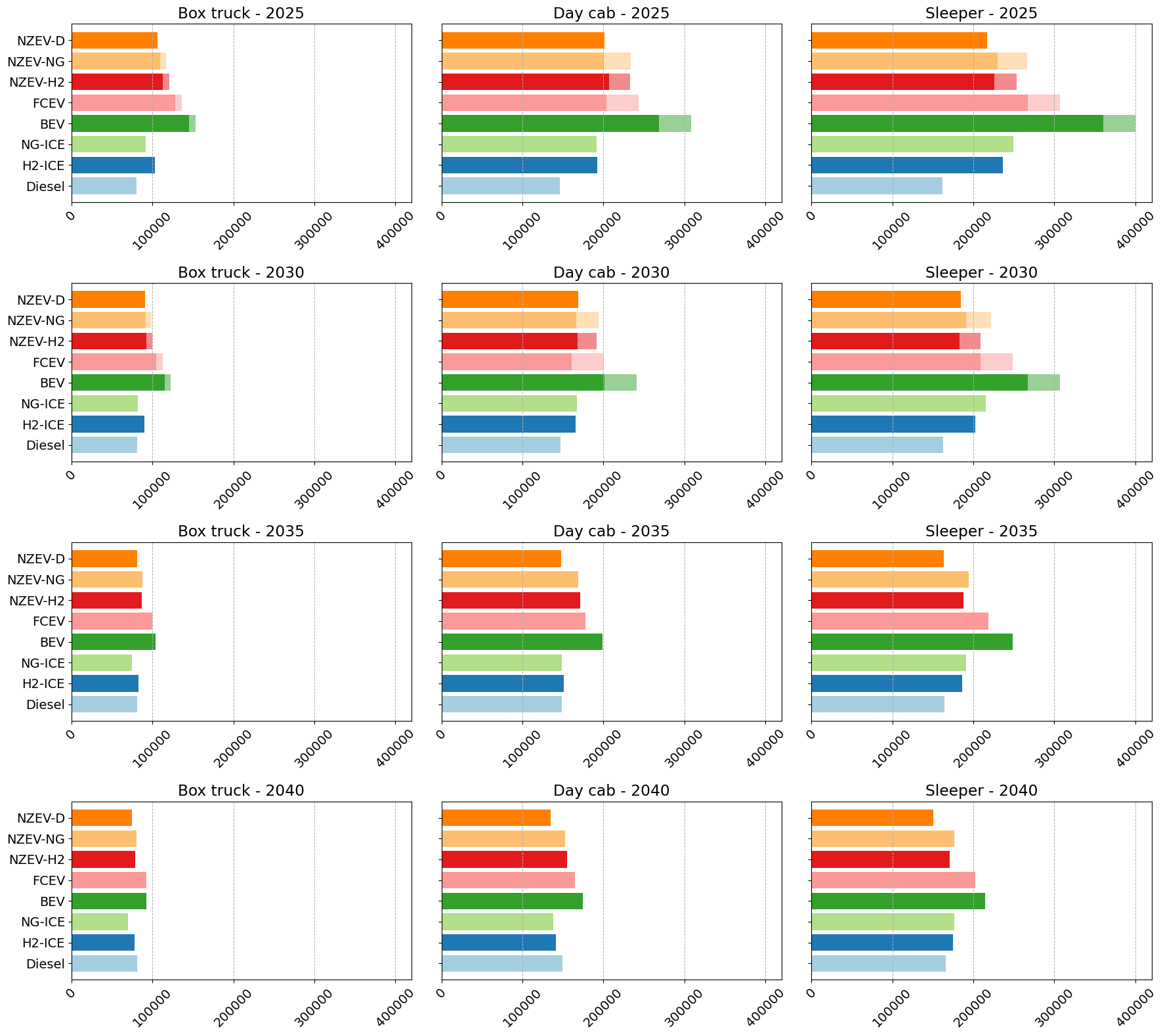}
     \caption{Vehicle price for four future years by three classes and eight powertrain types}
     \label{fig:vehicle_price4}
\end{figure}

We observe that diesel vehicles have the lowest prices, approximately \$100,000 for a box truck, \$140,000 for a day cab, and \$175,000 for a sleeper. Among the zero-emission vehicles (ZEV) and near-zero-emission vehicles (NZEV), fuel cell electric vehicles (FCEV) have the highest prices, followed by battery electric vehicles (BEV), NZEV0NG, NZEV-H2, NZEV-D, NG-ICE, and H2-ICE. It is also noted that the error bars for FCEVs are the largest, followed by BEVs and H2-ICEs, which is consistent with the significant variability in fuel cell and battery costs.

\begin{figure}[hbt!] 
\centering   
\includegraphics[width=\columnwidth]{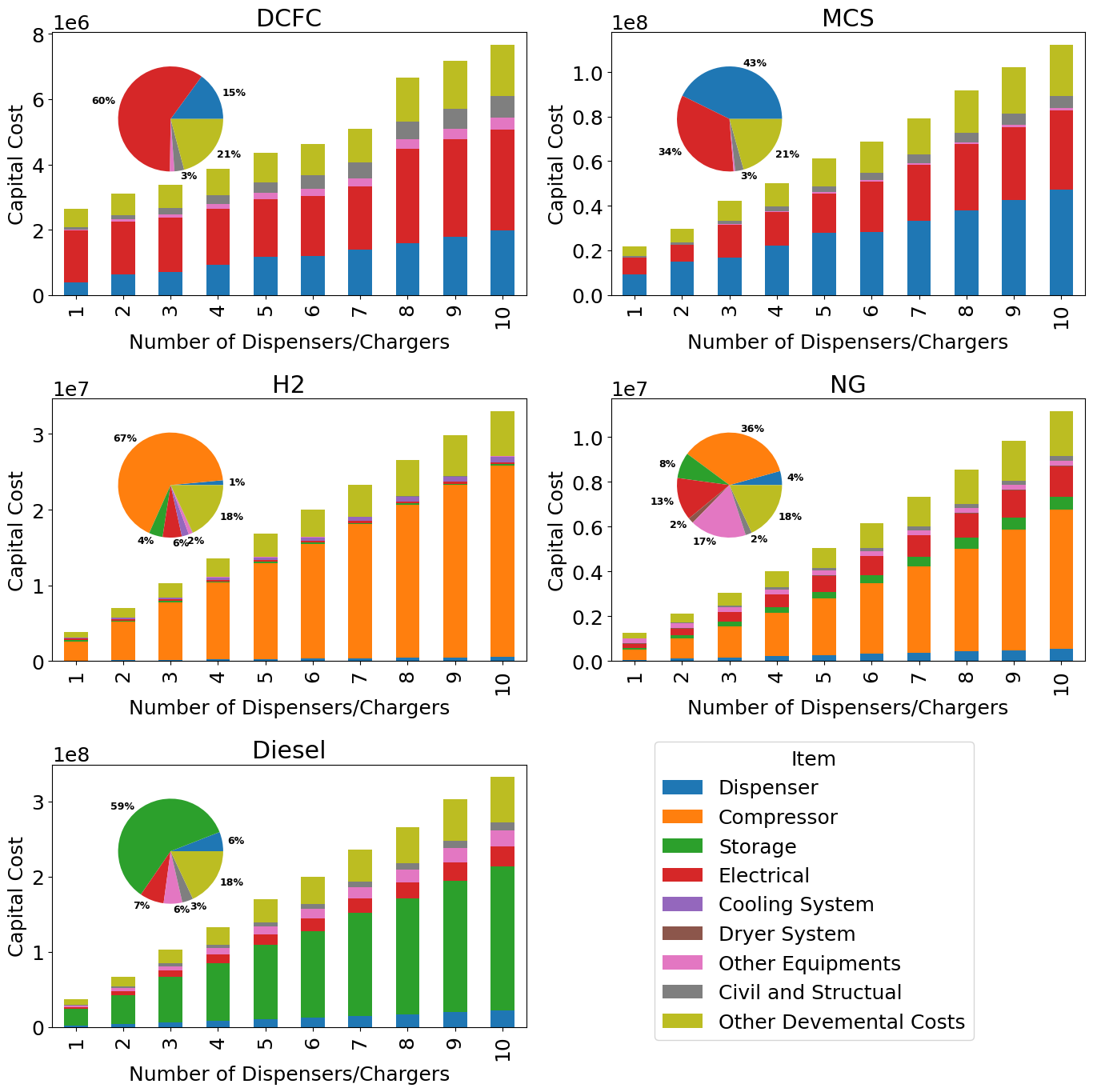}
     \caption{Capital Costs for Four Energy Types and Levelized CapEx for Five Infrastructures}
     \label{fig:cap_inf}
\end{figure}

We also employed the estimated technology learning curve from the ICCT report \citep{xie2023purchase} to calculate the technology learning rates for the key zero-emission powertrain components. The learning rate for batteries is estimated to be around 8\%, while the learning rates for fuel cells, hydrogen tanks, and electric drive trains are 9\%, 7\%, and 4\% respectively. Using these learning rates, we can project the impacts of technological evolution on vehicle prices.

Additionally, the IRA incentives on clean trucks, which were issued in 2023 and expire in 2033, are considered in the vehicle CapEx calculation.

Figure \ref{fig:vehicle_price4} presents the vehicle prices for four future years by three classes and eight powertrain types. The light shade represents the IRA incentives on specific types of trucks. It is observed that ZEVs and NZEVs become progressively cheaper in the future due to technological advancements, whereas diesel prices remain flat due to the maturity of diesel vehicle technology. By 2040, all NZEVs and ZEVs are projected to be cheaper than or equal to the price of diesel vehicles. BEVs, in particular, are expected to achieve cost competitiveness with diesel vehicles for box trucks and day cabs by 2035.

\subsubsection{Capital Costs of Infrastructures}

Infrastructure capital costs are composed of equipment costs and developmental costs. Equipment costs include expenses for dispensers/chargers, compressors, storage, cooling systems, dryer systems, electrical equipment, and other software, safety, and control equipment. Developmental costs encompass civil and structural costs as well as other related expenses. Figure \ref{fig:cap_inf} illustrates the capital costs for four energy types: diesel, hydrogen, electricity, and natural gas. We provide two options for charging infrastructure: DC Fast Charging (DCFC) and Megawatt Charging System (MCS). The key inputs for calculating capital costs are the charging power/filling rate and the number of dispensers/chargers. We assume the power of DCFC is 150 kW, MCS is 1250 kW, the filling rate for hydrogen is 1.8 kg/min, for natural gas is 70 gal/hour, and for diesel is 35 gal/min.

In Figure \ref{fig:cap_inf}, for each of the five infrastructures, the capital cost is shown with a detailed breakdown as the number of dispensers/chargers increases from 1 to 10. The figure also presents the percentage breakdown of these capital expenditures (CapEx) in a pie chart. We observe that electrical supply equipment constitutes 60\% of the DCFC CapEx and 15\% for dispensers. For MCS, the charger accounts for 43\% and the electrical supply equipment 34\%. Compressors are the major component of H2 and NG infrastructure CapEx, with 67\% and 36\% respectively. Storage represents about 59\% of the CapEx for diesel infrastructure. Developmental costs generally account for around 18\% of the capital cost for H2, NG, and diesel refueling infrastructure, and about 21\% for DCFC and MCS infrastructures.

\subsection{Vehicle TCO Results} 
To compare and understand the costs of different types of vehicles, our model calculates the TCO for each vehicle across the aforementioned eight categories, each providing a detailed cost breakdown.

\begin{table}[h!]
\centering
\caption{Variables and Assumptions}
\begin{tabular}{|p{3cm}|p{3cm}|p{10cm}|}
\hline
\textbf{Variables} & \textbf{Assumptions} & \textbf{Reference} \\ \hline
Discount rate & 7\% & \cite{hao2020range} \\ \hline
Interest rate & 4\% & Financial market statistics (2022) \\ \hline
Loan term & 5 years & Alonso-Villar et al. (2022);  Bhosale et al. (2022); and Rial and Pérez (2021) \\ \hline
Down payment & 20\% & Mansour and Haddad (2017); and Rial and Pérez (2021) \\ \hline
Vehicle useful lifetime & 5 years & An interview with a fleet organization \\ \hline
\end{tabular}
\label{table:assumptions}
\end{table}

\begin{figure}[htb] \centering   \includegraphics[width=\columnwidth]{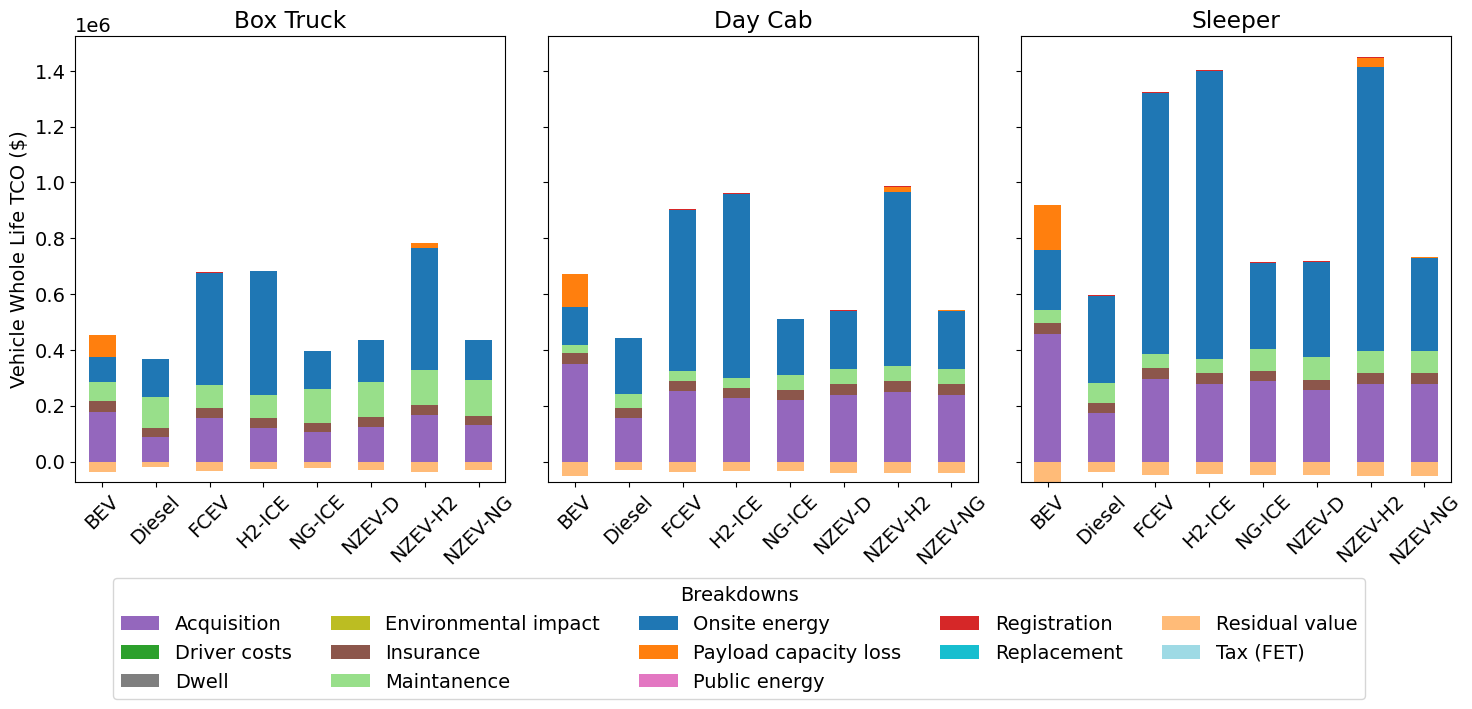}
     \caption{Vehicle whole life (5-years) TCO by three classes and eight powertrain types}
     \label{fig:vehicle_wtco}
\end{figure}

\begin{figure}[hbt!] \centering   \includegraphics[width=\columnwidth]{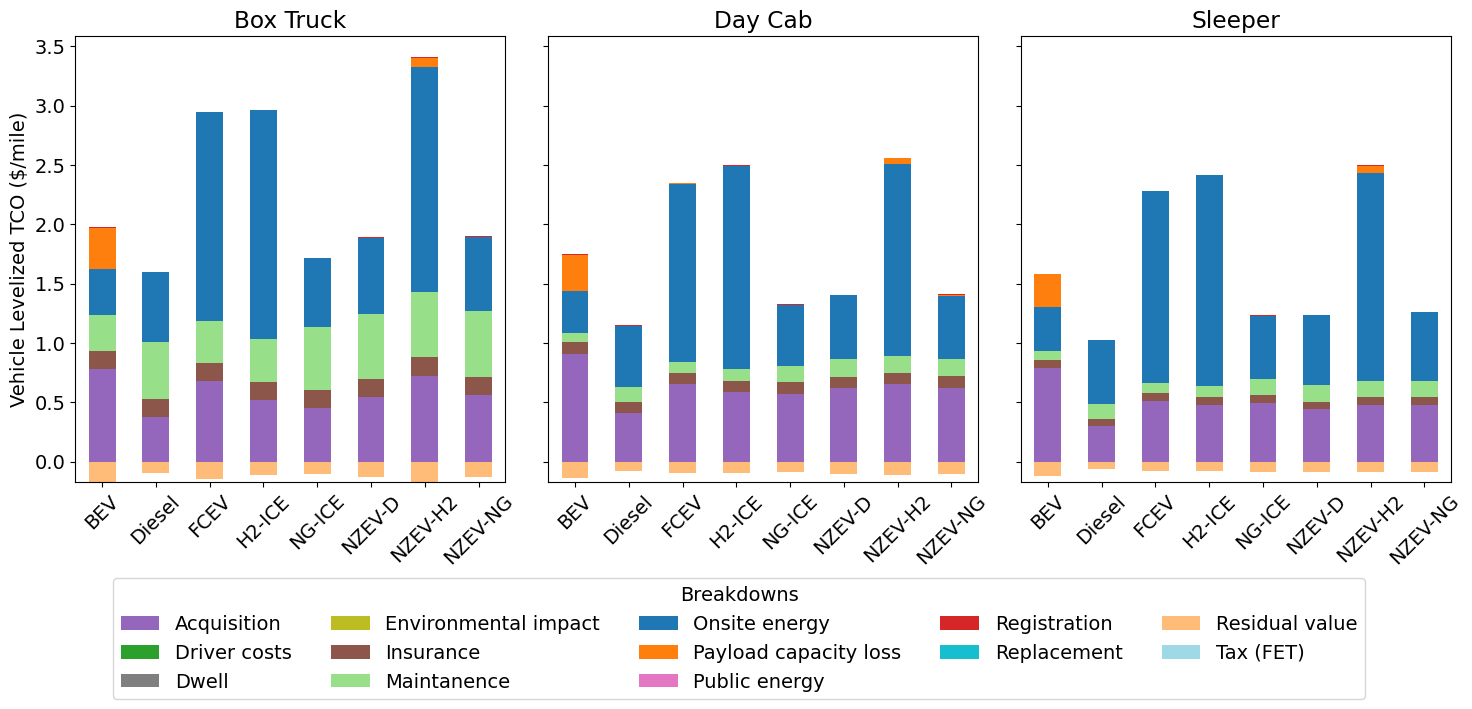}
     \caption{Vehicle Levelized TCO by three classes and eight powertrain types}
     \label{fig:vehicle_ltco}
\end{figure}

The financial assumptions, including the discount rate, interest rate, loan term, down payment, and vehicle useful life, are presented in Table \ref{table:assumptions}. Figures \ref{fig:vehicle_wtco} and \ref{fig:vehicle_ltco} illustrate the calculated vehicle whole life TCO and vehicle levelized TCO in 2023\$, respectively. The whole life cost for sleepers is higher than for day cabs and box trucks, with box trucks ranging from \$500,000 to \$1,000,000, day cabs from \$750,000 to \$1,600,000, and sleepers from \$1,150,000 to \$2,000,000.  However, the levelized cost for sleepers is lower than for day cabs, which in turn is lower than for box trucks. Specifically, the levelized cost for box trucks ranges from \$2.8/mile to \$5/mile, for day cabs from \$2.4/mile to \$4.7/mile, and for sleepers from \$1.8/mile to \$3.4/mile. This difference is due to the higher utilization rates of sleepers, resulting in greater VMT compared to day cabs and box trucks, which have lower VMT.

\begin{figure}[hbt!] \centering   \includegraphics[width=\columnwidth]{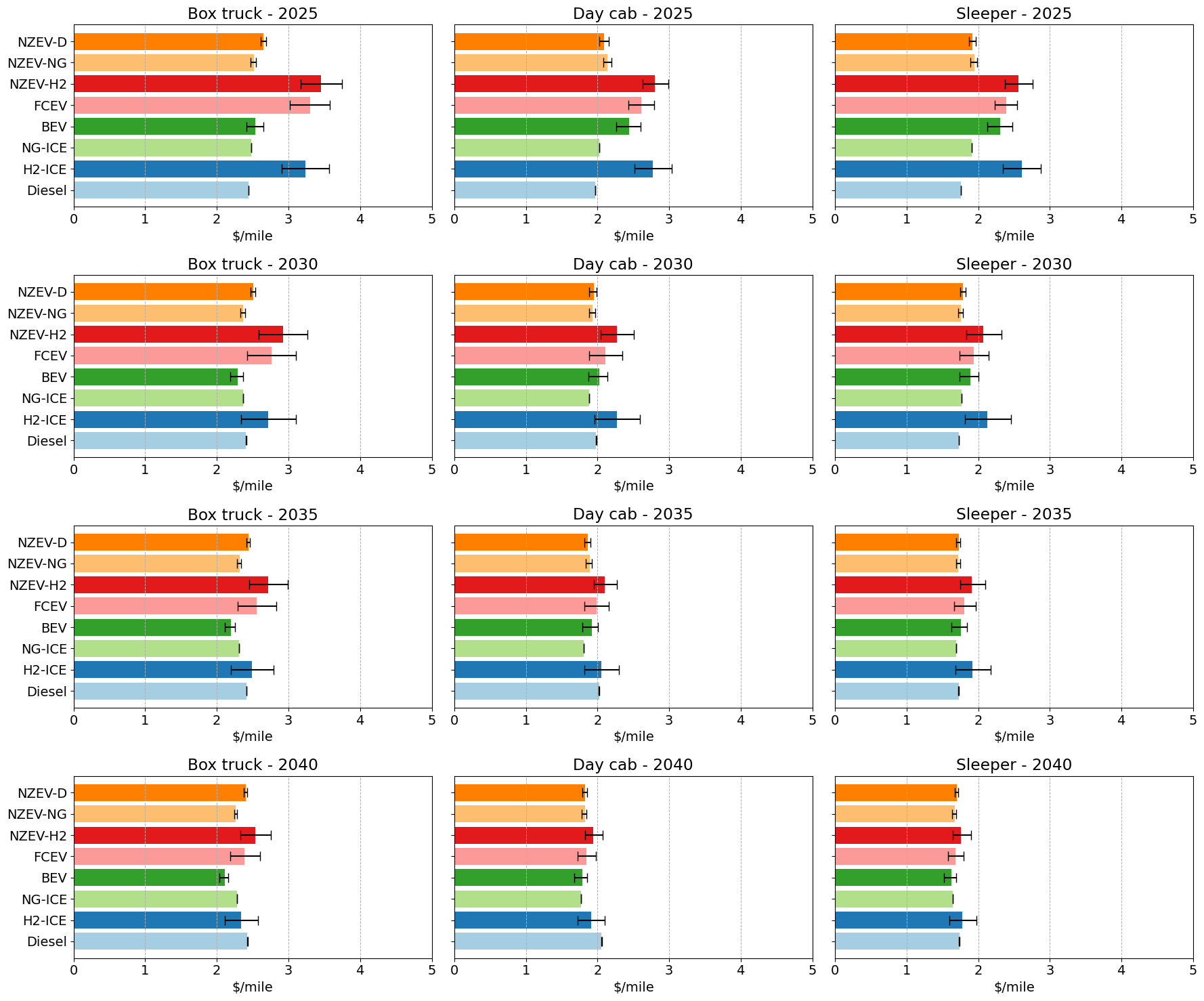}
     \caption{Vehicle Levelized TCO for 4 years by three classes and eight powertrain types}
     \label{fig:vehicle_tco4}
\end{figure}

It is observed that energy cost, driver cost, and acquisition cost are the top cost components for all vehicles. FCEVs, H2-ICEs, and NZEV-H2s have higher energy costs than other powertrain types due to the high price of hydrogen compared to the relatively lower prices of diesel, natural gas, and electricity. BEVs have the highest dwell time cost but the lowest maintenance cost among the powertrain types. Diesel vehicles have the highest environmental impact cost, although this has a minor impact on the overall TCO.

Considering technology advancements, future energy price projections from the \cite{EIA2023}, and other time-dependent assumptions, the model is structured to estimate the future TCO for all vehicle types under three levels of technological advancement: low, moderate, and high. Fig. \ref{fig:vehicle_tco4} illustrates the vehicle levelized TCO in \$/mile for the years 2025, 2030, 2035, and 2040, categorized by vehicle classes and powertrain types, under moderate technology advancement assumptions. The error bars represent the range across all three assumption scenarios. As shown in Fig. \ref{fig:vehicle_tco4}, we observe that vehicle TCO decreases over time, with diminishing gaps among different powertrain types. Notably, despite the higher capital costs of battery electric vehicles (BEVs) compared to diesels, the TCO for box trucks and day cab BEVs is projected to be lower than that of their diesel counterparts by around 2030, and for sleeper cab BEVs by around 2035. This earlier TCO parity is attributed to advancements in battery technology, lower maintenance costs, and reduced dwell costs, assuming the development of high-power infrastructure in the future.

Additionally, NG-ICE vehicles are expected to be competitive with diesel by 2030. The TCO for FCEVs and H2-ICE vehicles is anticipated to be lower than diesel by 2035. NZEVs are projected to reach TCO parity with diesel around 2040. This is because NZEVs combine conventional vehicle components with net zero emission vehicle technology, resulting in higher initial costs.


\subsection{System of system TCO}
This section presents the system of systems TCO by integrating infrastructure TCO into vehicle TCO as one of the cost components. This integration allows for a comprehensive comparison of diesel vehicles with alternative vehicles at the system level, considering both vehicle costs and the associated charging/refueling costs during operation. This comparison can provide valuable insights for fleet agencies when making investment decisions on alternative vehicles that can substitute diesel vehicles, considering the installation of on-site infrastructure to refuel and charge their fleet economically and meet CO2 emission reduction targets (e.g., ACF regulations). These decisions take into account time-sensitive factors such as technology progress in vehicles and infrastructure, energy prices, carbon prices, and policies and regulations.

Our analysis assumes a fleet of 30 day cab trucks that require overnight refueling or charging at the depot. We utilize real-world fleet movement data—including route information, VMT, and vehicle weight—provided by one of our interviewed experts. To investigate the system-of-systems cost for this fleet across all powertrain types and detailed variants of BEVs and FCEVs, we consider various BEV options with different battery sizes and FCEV options with different fuel cell power capacities. Thus, we configure 12 vehicle types: D-ICE, H2-ICE, NG-ICE, BEV500, BEV700, BEV1000, FCEV200, FCEV300, FCEV400, NZEV-D, NZEV-H2, NZEV-NG. In this configuration, BEV500 represents a BEV with a 500 kWh battery, BEV700 and BEV1000 represent BEVs with 700 kWh and 1000 kWh batteries, respectively. FCEV200 denotes an FCEV with 200 kW fuel cell power. The configuration for other vehicle types is consistent with Table \ref{table:powertrain_specs}. 

We conduct a what-if analysis for the fleet, assuming that all vehicles are of the same type selected from the 12 vehicle types. It is assumed that the fleet agency will install the corresponding infrastructure to supply energy for the fleet of a specific vehicle type under each scenario. For example, a diesel infrastructure will be developed if all trucks are diesel vehicles, and charging infrastructure (assuming DCFC 150kW) will be invested in if all trucks are BEVs. Similarly, hydrogen refueling infrastructure will be installed for FCEVs and H2-ICEs, and natural gas refueling infrastructure for NG-ICEs.
This problem can be modeled by considering a set of vehicles, indexed by \( i \), and a set of dispensers/chargers, indexed by \( j \). Each vehicle \( i \) arrives at the depot at time \( A_i \) and has a dwell time \( T_i \), representing the maximum time available for refueling/charging. The required charging time for each vehicle is denoted by \( R_i \). Each dispenser/charger \( j \) has a maximum operational time \( M_j \), typically 24 hours. The decision variables include \( x_{ij} \in \{0,1\} \), a binary variable indicating whether vehicle \( i \) is assigned to dispenser/charger \( j \); \( y_j \in \{0,1\} \), a binary variable indicating whether dispenser/charger \( j \) is utilized; and \( s_i \geq 0 \), the start time of charging for vehicle \( i \). Our objective is to determine the minimum number of dispensers/chargers required (\( N_c \)) and to devise a charging schedule that ensures all vebicles are fully refueled/charged within their dwell times and the operational constraints of the dispensers/chargers.

The optimization problem is formulated as follows:

\begin{align*}
\text{Minimize} \quad & N_c = \sum_{j} y_j \\
\text{subject to} \quad & \sum_{j} x_{ij} = 1, \quad \forall i \\
& x_{ij} \leq y_j, \quad \forall i, \forall j \\
& s_i + R_i \leq A_i + T_i, \quad \forall i \\
& s_i + R_i \leq s_k + M_j (1 - x_{ij} x_{kj}), \quad \forall i \neq k, \forall j \\
& W_{ij} + R_i \leq \min(T_i, M_j - A_i), \quad \forall i, \forall j \\
& s_i + R_i \leq M_j, \quad \forall i, \forall j \\
& x_{ij} \in \{0,1\}, \quad y_j \in \{0,1\}, \quad s_i \geq 0, \quad N_c \in \mathbb{Z}^+
\end{align*}

In this formulation, the first constraint ensures that each vehicle is assigned to exactly one dispenser/charger. The second constraint states that a dispenser/charger is considered utilized if at least one vehicle is assigned to it. The third constraint guarantees that the refueling of each vehicle is completed within its dwell time. The fourth constraint prevents overlap in refueling/charging schedules for vehicles assigned to the same charger by ensuring that their refueling/charging times do not overlap on the same dispenser/charger. The fifth constraint accounts for possible waiting times for vehicles assigned to dispensers/chargers, ensuring that the waiting time plus the required fueling time does not exceed the minimum of the dwell time and the dispenser/charger's remaining operational time. The sixth constraint ensures that no dispenser/charger operates beyond its maximum operational time. Finally, the last line defines the nature of the decision variables, specifying which are binary, continuous, or integer.

By solving this optimization problem, we determine the optimal number of chargers required as described in Table \ref{table:infrastructure_utilization}. We assume a construction period of two years, but when only one dispenser is needed, construction is completed in one year. Other assumptions include a system lifetime of 30 years \citep{saleh2008durability} and the lifetime of key depreciable equipment at 10 years \citep{saleh2008durability}, requiring replacement every 10 years.


\begin{table}[h!]
\centering
\caption{Infrastructure Utilization and Service Capacity}
\label{table:infrastructure_utilization}
\begin{tabular}{|c|c|c|c|}
\hline
\textbf{Powertrain} & \textbf{Number of Dispensers} & \textbf{Number of Chargers} & \textbf{Utilization Rate} \\
\hline
Diesel-ICE & 1 & 0 & 15\% \\
H2-ICE & 1 & 0 & 18\% \\
NG-ICE & 1 & 0 & 24\% \\
BEV1 & 0 & 7 & 18\% \\
BEV2 & 0 & 9 & 78\% \\
BEV3 & 0 & 11 & 82\% \\
FCEV1 & 1 & 0 & 20\% \\
FCEV2 & 1 & 0 & 20\% \\
FCEV3 & 1 & 0 & 20\% \\
NZEV-H2 & 1 & 3 & 9\% / 12\% \\
NZEV-NG & 1 & 3 & 12\% / 12\% \\
NZEV-D & 1 & 3 & 7\% / 12\% \\
\hline
\end{tabular}
\end{table}



\begin{figure}[hbt!] 
     \centering   \includegraphics[width=\columnwidth]{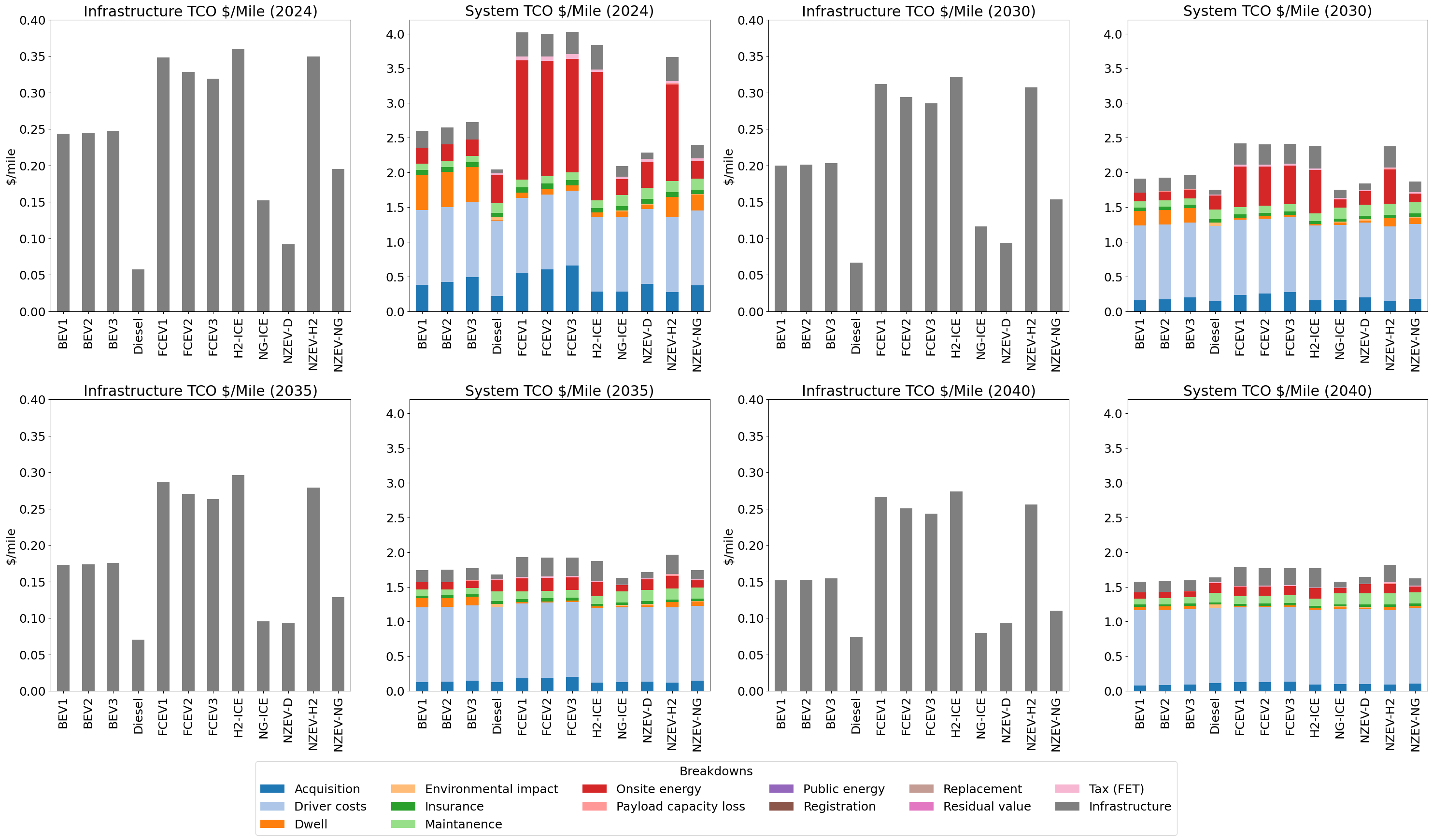}
     \caption{Vehicle average Levelized TCO with and without infrastructure costs for the fleet}
     \label{fig:veh_tco_inf_tco}
\end{figure}

Figure \ref{fig:veh_tco_inf_tco} shows the average vehicle levelized TCO in \$/mile, both with and without the infrastructure cost, for the fleet across four years of interest. The vehicle TCO inclusive of infrastructure cost is referred to as the system of system TCO in this study.

In 2023, the infrastructure cost per mile is approximately \$0.25 for BEVs, \$0.05 for diesel vehicles, \$0.30-\$0.35 for FCEVs, \$0.36 for H2-ICEs, \$0.14 for NG-ICEs, \$0.08 for NZEV-Ds, \$0.35 for NZEV-H2s, and \$0.20 for NZEV-NGs. Infrastructure costs for alternative fuels are expected to diminish over time due to technological advancements, while diesel infrastructure costs may increase slightly due to factors such as regulatory changes and market demand shifts \citep{rahimi2024beyond}. 

It is observed that the infrastructure cost significantly impacts the vehicle TCO. For instance, the infrastructure cost for BEVs is projected to decrease from around \$0.25/mile in 2023 to approximately \$0.15/mile by 2040, while the infrastructure cost for diesel vehicles remains around \$0.05/mile, potentially rising to \$0.06/mile by 2040. Including the amortized infrastructure cost in the vehicle TCO affects the breakeven timeline for alternative vehicles compared to their diesel counterparts. The analysis shows that BEVs will achieve cost parity with diesel vehicles by 2024, compared to 2035 without considering infrastructure costs. FCEVs and NZEV-H2s, however, will not match diesel vehicles before 2040 due to the higher incremental costs associated with hydrogen refueling infrastructure. NG-ICEs are expected to achieve cost parity with diesel vehicles by 2030, and NZEV-Ds and NZEV-NGs by 2035.

It is important to note that the system of system vehicle TCO results are highly dependent on the given assumptions. Changing assumptions, such as the technology learning rate for vehicles and equipment, will alter the results. Therefore, in the next section, we investigate the sensitivity of key assumptions and conduct scenario analyses to better understand the factors impacting costs and identify which factors can be adjusted to achieve cost-effectiveness.

\begin{figure}[hbt!] 
     \centering   \includegraphics[width=\columnwidth]{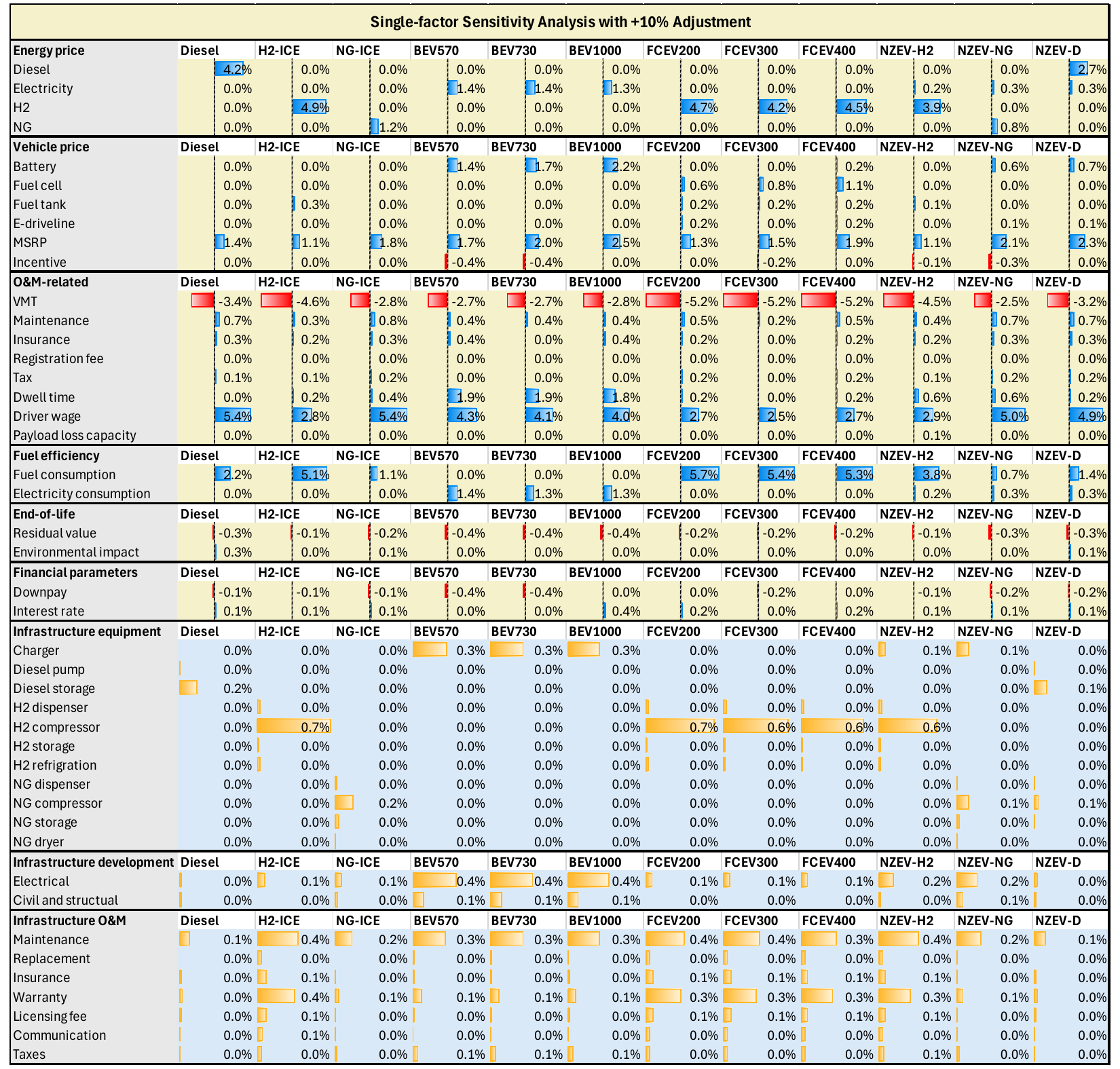}
     \caption{Single factor sensitivity analysis}
     \label{fig:sensitivity}
\end{figure}


\subsection{Sensitivity Analysis}

Figure \ref{fig:sensitivity} details the single-factor sensitivity analysis of various factors on the system of system TCO for the 12 vehicle types. We consider nine categories of factors: energy price, vehicle price, vehicle Operating and Maintenance (O\&M)-related factors, fuel efficiency, end-of-life, financial parameters, infrastructure equipment price, infrastructure development cost, and infrastructure O\&M-related factors. The sensitivity analysis involves adjusting each factor by +10\% to observe its impact on the system TCO.

The results show that energy price significantly affects the TCO for several vehicle types. For instance, a 10\% increase in diesel price results in a 4.2\% increase in diesel TCO and a 2.7\% increase in NZEV-D TCO. A 10\% increase in hydrogen price results in a 4.9\% increase in TCO for H2-ICEs and a 4.7\% increase for FCEV200s. Conversely, natural gas and electricity prices have less impact on non-natural gas and non-electric vehicles, respectively.

Vehicle price components, such as battery, fuel cell, and MSRP, notably impact the TCO. For BEV1000s, a 10\% increase in battery price results in a 2.2\% increase in TCO. Similarly, for FCEV400s, a 10\% increase in fuel cell price results in a 1.1\% increase in TCO.

Vehicle O\&M costs, such as vehicle miles traveled (VMT) and driver cost, also highly influence TCO. A 10\% increase in VMT decreases the TCO by 5.2\% for FCEV200s, highlighting the sensitivity of hydrogen fuel cell vehicles to utilization rates. A 10\% increase in driver wage increases the TCO by around 2.5\% to 5.4\%. Maintenance costs have a relatively lower impact, with a 10\% increase resulting in about a 0.8\% increase in TCO for NG-ICEs.

Fuel efficiency changes significantly impact the TCO of all vehicle types. For example, a 10\% increase in fuel consumption leads to a 5.7\% increase in TCO for FCEV200s and a 5.1\% increase for H2-ICEs. A 10\% increase in electricity consumption leads to a 1.3-1.4\% increase in TCO for BEVs. This sensitivity underscores the importance of fuel efficiency in managing TCO.

End-of-life factors, such as residual value and environmental impact costs, have smaller effects on TCO. A 10\% increase in residual value reduces the TCO by about 0.4\% for BEV570s and 0.3\% for diesel vehicles. Environmental impact costs, though generally lower, can still influence the overall TCO.

Financial assumptions like down payment and interest rates also affect TCO. A 10\% increase in the down payment reduces the TCO by 0.4\% for BEV570s, while a similar increase in interest rate has a negligible effect on most vehicle types.

Infrastructure equipment prices, such as chargers and compressors, particularly affect BEVs and hydrogen vehicles. For instance, a 10\% increase in the cost of hydrogen compressors raises the TCO by 0.7\% for H2-ICEs and 0.6\% for FCEV200s. Development costs for infrastructure, including electrical and civil/structural costs, have varied impacts. A 10\% increase in electrical infrastructure costs results in a 0.4\% increase in TCO for BEV570s, indicating the significance of infrastructure development costs. 

O\&M costs for infrastructure, such as maintenance and insurance, also affect TCO. A 10\% increase in maintenance costs for infrastructure raises the TCO by about 0.4\% for FCEV200s and 0.3\% for BEV570s, demonstrating the sensitivity of these factors.

The single-factor sensitivity analysis highlights that energy price, vehicle price, and fuel efficiency are the most significant factors influencing TCO across different vehicle types. These insights can guide fleet agencies in making informed decisions when considering alternative vehicles and infrastructure investments.

\section{Conclusion} \label{conclusion}
This study developed a comprehensive TCO model to evaluate the economic viability and strategic pathways for decarbonizing road freight transport. The model integrates three interconnected modules—vehicle and infrastructure—to provide a holistic assessment of alternative fuel technologies, charging and refueling infrastructure requirements, and the integration of power grid support with renewable energy generation. We focused on medium- and heavy-duty vehicle classes, encompassing eight powertrain types, and evaluated the costs associated with vehicle acquisition, operation, maintenance, energy consumption, environmental impacts, and end-of-life considerations. By incorporating detailed financial parameters, technological advancements, and policy incentives, the model offers valuable insights for fleet agencies considering the transition to alternative fuel vehicles.

Our findings indicate that the costs for ZEVs and NZEVs are currently higher than those for diesel vehicles. However, the projected declines in battery, fuel cell, and hydrogen tank costs, coupled with advancements in electric drive technologies and policy incentives, are expected to significantly reduce these costs over time. 

Despite the robustness of our model, several limitations should be acknowledged. First, the reliance on various assumptions and projections, such as energy prices, technological learning rates, and policy incentives, which may introduce uncertainties in the cost estimations. Future research will focus on fully developing the "regional" infrastructure module and incorporating revenue monetization for shared infrastructure utilization. This enhancement will provide a more comprehensive assessment of the economic implications of decarbonizing freight transport. Additionally, we plan to conduct a comprehensive vehicle parity analysis under various conditions, extending the analysis to regional-specific and national scales. This future work will offer deeper insights into the strategic pathways for achieving cost-effective and sustainable freight transport solutions.

In conclusion, this study underscores the importance of considering both vehicle and infrastructure costs in evaluating the economic viability of alternative fuel vehicles. Our findings provide a roadmap for fleet agencies and policymakers to make informed decisions, facilitating the transition to a more sustainable and economically viable road freight transport system.









\bibliographystyle{elsarticle-harv} 
\bibliography{ref}

\end{document}